\documentstyle{mn}
\input epsf
\epsfverbosetrue
\newcommand{\bell}{\mbox{\boldmath$\ell$}}
\begin{document}
\title[Vertical shear flow and turbulence]{The response of a turbulent 
accretion disc to an imposed epicyclic shearing motion}
\author[U. Torkelsson et al.]{Ulf Torkelsson$^{1,2}$, Gordon I. 
Ogilvie$^{1,3,4}$,
Axel Brandenburg$^{3,5,6}$, James E. Pringle$^{1,3}$, \newauthor
\AA ke Nordlund$^{7,8}$, Robert F.
Stein$^9$\\
$^1$Institute of Astronomy, Madingley Road, Cambridge CB3 0HA, United Kingdom \\
$^2$Chalmers University of Technology/G\"oteborg University, Department of
Theoretical Physics, Astrophysics Group, \\
S-412 96 Gothenburg, Sweden\\
$^3$Isaac Newton Institute for Mathematical Sciences, 20 Clarkson Road, 
Cambridge CB3 0EH, United Kingdom\\
$^4$Max-Planck-Institut f\"ur Astrophysik, Karl-Schwarzschild-Stra\ss e 1,
Postfach 1523, D-85740 Garching bei M\"unchen, Germany\\
$^5$Department of Mathematics, University of Newcastle upon Tyne, NE1 7RU,
United Kingdom\\
$^6$Nordita, Blegdamsvej 17, DK-2100 Copenhagen \O, Denmark\\ 
$^7$Theoretical Astrophysics Center, Juliane Maries Vej 30, DK-2100 Copenhagen
\O, Denmark\\
$^8$Copenhagen University Observatory, Juliane Maries Vej 30, DK-2100 Copenhagen
\O, Denmark\\
$^9$Department of Physics and Astronomy, Michigan State University, East
Lansing, MI 48824, USA}
\maketitle

\begin{abstract}
We excite an epicyclic motion, whose amplitude depends on the vertical
position, $z$, in a simulation of a turbulent accretion disc.  An
epicyclic motion of this kind may be caused by a warping of the disc.
By studying how the epicyclic motion decays 
we can obtain information about the interaction between the warp and the
disc turbulence.  A high amplitude epicyclic motion decays first by
exciting inertial waves through a parametric instability, but its
subsequent exponential damping may be reproduced by a turbulent
viscosity.
We estimate the effective viscosity parameter,
$\alpha_{\rm v}$, pertaining to such a vertical shear.  We also gain
new information on the properties of the disc turbulence in general,
and measure the usual viscosity parameter, $\alpha_{\rm h}$,
pertaining to a horizontal (Keplerian) shear.  We find that, as is
often assumed in theoretical studies, $\alpha_{\rm v}$ is
approximately equal to $\alpha_{\rm h}$ and both are much less than
unity, for the field strengths achieved in our local box calculations
of turbulence.  
In view of the smallness ($\sim 0.01$) of $\alpha_{\rm v}$ and 
$\alpha_{\rm h}$ we conclude that for $\beta = p_{\rm gas}/p_{\rm mag}
\sim 10$ the timescale for diffusion or damping of a
warp is much shorter than the usual viscous timescale.  Finally, we
review the astrophysical implications.
\end{abstract}

\section{Introduction}

Warped accretion discs appear in many astrophysical systems.  A well
known case is the X-ray binary Her X-1, in which a precessing warped
disc is understood to be periodically covering our line of sight to
the neutron star, resulting in a 35-day periodicity in the X-ray
emission (Tananbaum et al.
1972; Katz 1973; Roberts 1974).  A similar phenomenon is believed
to occur in a number of other X-ray binaries.  In recent years the
active galaxy NGC 4258 has received much attention as a warp in the
accretion disc has been made visible by a maser source (Miyoshi et
al. 1995).

A warp may appear in an accretion disc in response to an external
perturber such as a binary companion, but it is also possible that the
disc may produce a warp on its own.  Pringle \shortcite{pringle}
showed that the radiation pressure from the central radiation source
may produce a warp in the outer disc.  In a related mechanism the
irradiation can drive an outflow from the disc.  The force of the wind
may then in a similar way excite a warp in the disc (Schandl \& Meyer
1994).

Locally, one of the effects of a warp is to induce an epicyclic motion
whose amplitude varies linearly with distance from the midplane of the
disc.  This motion is driven near resonance in a Keplerian disc, and
its amplitude and phase are critical in determining the evolution of
the warp (Papaloizou \& Pringle 1983; Papaloizou \& Lin 1995).  
Depending on the strength of
the dissipative process the warp may either behave as a propagating
bending wave or evolve diffusively.
In the latter case the 
amplitude of the epicyclic motion is determined by the dissipative
process.

A (possibly) related dissipative process is responsible for driving
the inflow and heating the disc by transporting
angular momentum outwards.  From a 
theoretical point of view this transport has
been described in terms of a viscosity \cite{shakura:sunyaev}, but the source 
of the viscosity remained uncertain for a long time.  It was clear from the
beginning that molecular viscosity would be insufficient, so one appealed to
some form of anomalous viscosity presumably produced by turbulence in the
accretion disc.  However the cause of the turbulence could not be found
as the Keplerian rotation is hydrodynamically stable according to Rayleigh's
criterion.

Eventually Balbus \& Hawley \shortcite{balbus:hawley} discovered that
the Keplerian flow becomes unstable in the presence of a magnetic field.
This magnetic shearing instability had already been described by 
Velikhov \shortcite{velikhov} and Chandrasekhar \shortcite{chandrasekhar},
but it had not been thought applicable in the context of accretion discs
before.
Several numerical simulations (e.g. Hawley, Gammie \& Balbus
1995, Matsumoto \& Tajima 1995, Brandenburg et al. 1995, Stone et
al. 1996) have demonstrated 
how this instability generates turbulence in a Keplerian shear flow.  

The most
important result of these simulations has been to demonstrate that the
Maxwell and Reynolds stresses that the turbulence generates will
transport angular momentum outwards, thus
driving the accretion.  The energy source of the turbulence is the Keplerian
shear flow, from which the magnetic field taps energy.  This energy is then
partially dissipated due to Ohmic diffusion, but an equal amount of energy
is spent on exciting the turbulent motions.
The turbulent motions on the other hand give rise to a dynamo, which sustains
the magnetic
field required by the Balbus-Hawley instability (Brandenburg et al.
1995, Hawley, Gammie \& Balbus 1996).

In general the energy of the magnetic field is an order of magnitude larger
than the energy of the turbulent velocities, but almost all of the magnetic
energy is associated with the toroidal magnetic field, and the poloidal 
magnetic field components are comparable to the turbulent velocities.

So far none of the simulations has addressed
the question of how the turbulence responds to external perturbations
or systematic motions that are more complex than a Keplerian shear
flow.  The purpose of this paper is to begin such an
investigation by studying how the turbulence interacts with an imposed
shearing epicyclic motion of the type found in a warped disc.

We start this paper by describing the shearing-box approximation of
magnetohydrodynamics and summarizing the properties of the epicyclic
motion in a shearing box in Sect. 2.  Section 3 is then a description
of our simulations of an epicyclic motion.  The results of the
simulations are then described in Sect. 4 and briefly summarized in
Sect. 5.

\section{Mathematical formulation}

\subsection{The local structure of a steady disc}

For the intentions of this paper it is sufficient to use a simple
model of the vertical structure of a geometrically thin accretion
disc.  The disc is initially in hydrostatic equilibrium,
\begin{equation}
  \frac{\partial p}{\partial z} = \rho g_z,
\end{equation}
where $p$ is the pressure, $\rho$ the density, and $g_z = -GMz/R_0^3$
the vertical component of the gravity with $G$ the gravitational
constant, $M$ the mass of the accreting star, and $R_0$ the radial
distance from the star.  For simplicity we assume that the disc
material is initially isothermal, and is a perfect gas, so that $p =
c_{\rm s}^2 \rho$, where $c_{\rm s}$ is the isothermal sound speed,
which is initially constant.  The density distribution is then
\begin{equation}
  \rho = \rho_0 \mbox{e}^{-z^2/H^2},
\end{equation} 
where the Gaussian scale height, $H$, is given by 
\begin{equation}
  H^2 = \frac{2c_{\rm s}^2R_0^3}{GM}.
\end{equation}

\subsection{Epicyclic motion in the shearing box approximation}

In the shearing box approximation a small part of the accretion disc
is represented by a Cartesian box which is rotating at the Keplerian
angular velocity $\Omega_0 = \sqrt{GM/R_0^3}$.  The box uses the
coordinates $(x,y,z)$ for the radial, azimuthal and vertical
directions, respectively.  The Keplerian shear flow within the box is
$u_y^{\left(0\right)} = - \frac{3} {2} \Omega_0 x$, and we solve for
the deviations from the shear flow exclusively.  The
magnetohydrodynamic (MHD) equations may then be written
\begin{equation}
  \frac{{\cal D}\rho}{{\cal D}t} = - \nabla \cdot\left(\rho {\bf u}\right),
\label{cont_eq}
\end{equation}
\begin{equation}
  \frac{{\cal D}{\bf u}}{{\cal D}t} = -\left({\bf u\cdot \nabla}\right){\bf u} + 
{\bf g} + {\bf f}\left({\bf u}\right)- \frac{1}{\rho}{\bf \nabla}p
+ \frac{1}{\rho}{\bf J\times B} + \frac{1}{\rho}{\bf \nabla \cdot}\left(
2\nu\rho {\bf S}\right),
\label{mom_eq}
\end{equation}
\begin{equation}
  \frac{{\cal D}{\bf B}}{{\cal D}t} = {\bf \nabla \times} 
\left({\bf u \times B}\right) - {\bf \nabla \times} \eta \mu_0 {\bf J},
\label{ind_eq}
\end{equation}
\begin{equation}
  \frac{{\cal D}e}{{\cal D}t} = -\left({\bf u\cdot \nabla}\right)e -
\frac{p}{\rho} {\bf \nabla \cdot u} + \frac{1}{\rho} {\bf \nabla} \cdot
\left(\chi \rho
{\bf \nabla}e\right) + 2 \nu {\bf S}^2 + \frac{\eta \mu_0}{\rho}{\bf J}^2 +Q,
\label{en_eq}
\end{equation}
where ${\cal D}/{\cal D}t = \partial/\partial t +u_y^{\left(0\right)}
\partial/\partial y$ includes the advection by the shear flow, $\rho$ is the 
density, ${\bf u}$ the deviation from the Keplerian shear flow, 
$p$ the pressure,
${\bf f}({\bf u}) = \Omega_0(2 u_y, -\frac{1}{2}u_x, 0)$
the inertial force, ${\bf B}$ the magnetic field, ${\bf J} = 
{\bf \nabla \times B}/\mu_0$ the current, $\mu_0$ the permeability of free 
space, $\nu$ the viscosity, $S_{ij} = \frac{1}{2}(u_{i,j} + u_{j,i} - 
\frac{2}{3}\delta_{ij}u_{k,k})$ the trace-free
rate of strain tensor, $\eta$ the magnetic
diffusivity, $e$ the internal energy, $\chi$ the thermal conductivity, and $Q$
is a cooling function.  
The radial component of the gravity cancels against the
centrifugal force, and the remaining vertical component is
${\bf g}= -\Omega_0^2 z {\bf \hat z}$.
We adopt the equation of state for an ideal gas, $p = (\gamma -1)\rho e$.

When the horizontal components of the momentum equation (\ref{mom_eq})
are averaged over horizontal layers (an operation denoted by angle
brackets), we obtain
\begin{equation}
  \frac{\partial}{\partial t}\langle \rho u_x\rangle = 2 \Omega_0 
\langle\rho
u_y\rangle -\frac{\partial}{\partial z} \langle \rho u_x u_z\rangle +
\frac{\partial}{\partial z}\left\langle \frac{B_x B_z}{\mu_0}\right\rangle,
\label{epicx2}
\end{equation}
\begin{equation}
  \frac{\partial}{\partial t} \langle \rho u_y\rangle = - \frac{1}{2} \Omega_0
\langle \rho u_x\rangle - \frac{\partial}{\partial z}\langle \rho u_y u_z\rangle
+ \frac{\partial}{\partial z}\left\langle \frac{B_y B_z}{\mu_0}\right\rangle.
\label{epicy2}
\end{equation}
The explicit viscosity, which is very small, has been neglected here.
These equations contain vertical derivatives of components of the
turbulent Reynolds and Maxwell stress tensors, distinct from the
$xy$-components that drive the accretion.

We initially neglect the turbulent stresses and obtain the solution
\begin{equation}
\langle\rho u_x\rangle = \rho_0(z)\tilde u_0\left(z\right) \cos(\Omega_0 t),
\label{u-x}
\end{equation}
\begin{equation}
\langle\rho u_y\rangle = -\frac{1}{2}\rho_0(z)\tilde u_0\left(z\right) \sin(\Omega_0 t),
\label{u-y}
\end{equation}
which describes an epicyclic motion.  Here $\rho_0(z)$ is the initial
density profile.  The initial velocity amplitude $\tilde u_0$ is an
arbitrary function of $z$.  For the simulations in this paper we will
take $\tilde u_0(z)\propto \sin (kz)$, where $k = \pi/L_z$, and $- \frac{1}{2}
L_z \le z \le \frac{1}{2} L_z$
is the vertical extent of our shearing box.  This velocity profile is
compatible with the stress-free boundary conditions that we employ in
our numerical simulations, and gives a fair representation of a linear
profile close to the midplane of the disc.

The kinetic energy of the epicyclic motion is not conserved,
but the square of the epicyclic momentum
\begin{equation}
  E\left(z, t\right) = \frac{1}{2} \left\langle \rho u_x\right\rangle^2
+2 \left\langle \rho u_y\right\rangle^2,
\end{equation}
is conserved in the absence of turbulent stresses.  By
multiplying Eq. (\ref{epicx2}) by $\langle\rho u_x\rangle$, and Eq.
(\ref{epicy2}) by $4 \langle \rho u_y\rangle$ we obtain
\begin{equation}
  \frac{\partial E}{\partial t} = F_u + F_B,
\end{equation}
where
\begin{equation}
F_u = - \left \langle\rho u_x \right \rangle 
\frac{\partial} {\partial z} \left\langle\rho u_x u_z\right\rangle -
4\left\langle \rho u_y\right \rangle \frac{\partial}{\partial z} 
\left \langle \rho u_y u_z \right\rangle
\end{equation}
and
\begin{equation}
F_B = \left \langle \rho u_x \right
\rangle \frac{\partial}{\partial z} \left\langle \frac{B_x B_z}{\mu_0}
\right \rangle + 4 \left \langle \rho u_y \right \rangle 
\frac{\partial}{\partial z} \left\langle \frac{B_y B_z}{\mu_0}\right \rangle
\end{equation}
represent the `rates of working' of the Reynolds and Maxwell stresses,
respectively, on the epicyclic oscillator.  We may expect that both
$F_u$ and $F_B$ are negative, but by measuring them in the simulation
we may determine the relative importance of the Reynolds and Maxwell
stresses in damping the epicyclic motion.  
We will also refer to an epicyclic velocity amplitude
\begin{equation}
\tilde u=\sqrt{\langle u_x\rangle^2+4\langle u_y\rangle^2}.
\label{epic_amp}
\end{equation}

\subsection{Theoretical expectations}

The detailed fluid dynamics of a warped accretion disc has been
discussed by, e.g., Papaloizou \& Pringle
\shortcite{papaloizou:pringle}, Papaloizou \& Lin
\shortcite{papaloizou:lin} and Ogilvie \shortcite{ogilvie}.  The
dominant motion is circular Keplerian motion, but the orbital plane
varies continuously with radius $r$ and time $t$.  This may
conveniently be described by the tilt vector $\bell(r,t)$, which is a
unit vector parallel to the local angular momentum of the disc
annulus at radius $r$.  A dimensionless measure of the amplitude of
the warp is then $A = |\partial \bell/\partial \ln r|$.

In the absence of a detailed understanding of the turbulent stresses
in an accretion disc, it is invariably assumed that the turbulence acts as
an isotropic effective viscosity in the sense of the Navier-Stokes
equation.  In such an approach the dynamic viscosity is often parametrized as
\begin{equation}
\mu=\alpha p/\Omega_0,\label{mu}
\end{equation}
where $\alpha$ is a dimensionless parameter (Shakura \& Sunyaev 1973).
Although it is now possible to simulate the local turbulence in an
accretion disc, this form of phenomenological description of the turbulent
stress
is still valuable as it is not yet possible to study simultaneously both the
small-scale turbulence and the global dynamics of the accretion disc in a
numerical simulation.
One of the goals of this paper is to test the validity of this
hypothesis by comparing the predictions of the viscous model, as summarized
below, with the results of the numerical model.
We generalize the viscosity prescription by allowing, in a simple way, for 
the possibility that
the effective viscosity is anisotropic (cf. Terquem 1998).  The
parameter $\alpha_{\rm h}$ pertaining to `horizontal' shear
(i.e. horizontal-horizontal components of the rate-of-strain tensor,
such as the Keplerian shear) may be different from the parameter
$\alpha_{\rm v}$ pertaining to `vertical' shear
(i.e. horizontal-vertical components of the rate-of-strain tensor,
such as the shearing epicyclic motion).

\begin{figure}
\epsfxsize=8.8cm
\epsfbox{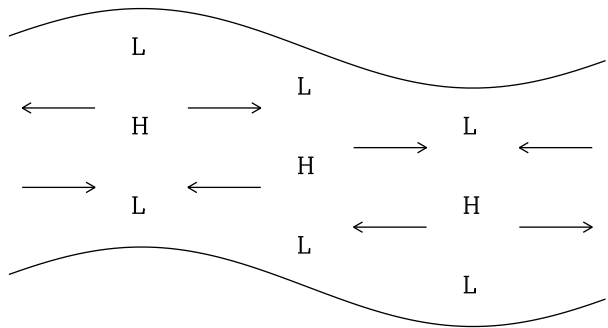}
\caption{Owing to the stratification there appear horizontal pressure gradients
in the warped disc.  These gradients excite the epicyclic motion (arrows
indicate forces, not velocities)}
\label{cartoon}
\end{figure}

Owing to the pressure stratification, resulting from the vertical
hydrostatic equilibrium, in a warped disc there are strong {\it
horizontal\/} pressure gradients (Fig. \ref{cartoon}), which generate
horizontal
accelerations of order $A\Omega_0^2 z$, that oscillate at the local orbital
frequency, as viewed in a frame co-rotating with the fluid.  
In a Keplerian disc the frequency of the horizontal pressure gradients
coincides with the natural
frequency of the resulting epicyclic motion, and a resonance occurs.
The amplitude of the resulting epicyclic motion depends on the amount of
dissipation present.  At low viscosities, $\alpha_{\rm v}\la H/r$, the
amplitude is limited by the coupling of the epicyclic motion to the 
vertical motion in a propagating bending wave, which transports energy
away.  At higher
viscosities
the amplitude is limited by a balance between the forcing and
the viscous dissipation, and the warp evolves diffusively
\cite{papaloizou:pringle}.
When $H/r \la \alpha_{\rm v} \ll 1$ the amplitude of
the epicyclic motion is
\begin{equation}
  u_x \propto u_y \propto \frac{A\Omega_0 z}{\alpha_{\rm v}}.
\label{epic_vel}
\end{equation}
The resulting hydrodynamic stresses $\overline{\rho u_x u_z}$ and
$\overline{\rho u_y u_z}$ (overbars denote averages over the orbital
timescale), which tend to flatten out the disc, are
also proportional to $\alpha_{\rm v}^{-1} A$ and therefore dominate
over the stresses $\propto \alpha_{\rm v}A$ due to small-scale turbulent
motions, which would
have the same effect.  
It is for this reason that the timescale for
flattening a warped disc is anomalously short compared to the usual
viscous timescale, by a factor of approximately $2 \alpha_{\rm h}
\alpha_{\rm v}$.  
(For more details, see Papaloizou
\& Pringle 1983 and Ogilvie 1999.  In these papers it was assumed that
$\alpha_{\rm h}=\alpha_{\rm v}$.)

We aim to measure both $\alpha_{\rm h}$ and $\alpha_{\rm v}$ in the
simulations.  The first may be obtained through the relation
\begin{equation}
\left\langle\rho u_x u_y-\frac{B_x
B_y}{\mu_0}\right\rangle_{\rm V}=\frac{3}{2}\alpha_{\rm h}\langle
p\rangle_{\rm V},
\label{alpha-h}
\end{equation}
which follows by identifying the total turbulent $xy$-stress with the
effective viscous $xy$-stress resulting from the viscosity (\ref{mu})
acting on the Keplerian shear (note that the definition of $\alpha_{\rm h}$
differs from that of $\alpha_{\rm SS}$ in Brandenburg et al. 1995 by a
factor $\sqrt{2}$).  Here the average is over the entire
computational volume.

The coefficient $\alpha_{\rm v}$ may be obtained by measuring the
damping time of the epicyclic motion.  In the shearing-box
approximation the horizontal components of the Navier-Stokes equation
for a free epicyclic motion decaying under the action of viscosity are
\begin{equation}
  \frac{\partial u_x}{\partial t} =
2 \Omega_0 u_y + \frac{1}{\rho} \frac{\partial}{\partial z}
\left(\frac{\alpha_{\rm v} p}{\Omega_0} \frac{\partial u_x}{\partial z}
\right),
\label{epicxg}
\end{equation}
\begin{equation}
  \frac{\partial u_y}{\partial t} = 
- \frac{1}{2} \Omega_0 u_x + \frac{1}{\rho} \frac{\partial}{\partial z}
\left( \frac{\alpha_{\rm v} p}{\Omega_0} \frac{\partial u_y}{\partial z}
\right).
\label{epicyg}
\end{equation}
Under the assumptions that $\alpha_{\rm v}$ and $\partial_z {\bf u}$
are independent of $z$ and that the
disc is vertically in hydrostatic equilibrium,
these equations have the exact solution
\begin{equation}
  u_x = C\Omega_0 z\,\mbox{e}^{-t/\tau}\cos(\Omega_0 t),
\end{equation}
\begin{equation}
  u_y = - \frac{1}{2} C \Omega_0 z\,\mbox{e}^{-t/\tau} \sin(\Omega_0 t),
\end{equation}
where $C$ is a dimensionless constant and
\begin{equation}
  \tau = \frac{1}{\alpha_{\rm v} \Omega_0}
\label{alpha-v}
\end{equation}
is the damping time.  
Admittedly it is already believed 
that $\alpha_{\rm h}$ is not independent of $z$
\cite{kyoto} and therefore $\alpha_{\rm v}$ may not
be either.  Also our velocity profile is not exactly proportional
to $z$.  However,
it is in fact the damping
time that matters for the application to a warped disc, and the solution that
we describe above is in a sense the fundamental mode of the epicyclic shear flow
in a warped disc.

It might be argued that the decaying epicyclic motion is fundamentally 
impossible in the presence of MHD turbulence.  After all, the
magnetorotational instability works because a magnetic coupling between   
two fluid elements executing an epicyclic motion allows an exchange of
angular momentum that destabilizes the motion.  However, 
we argue that the epicyclic motion must be {\it
re-stabilized}\/ in the nonlinear turbulent state.  Otherwise epicyclic
motions, which are continuously and randomly forced by the turbulence,   
would grow indefinitely \cite{bh-rev}.  
In fact they last only a few orbits, as
discussed in Section 3.2 below.  Moreover, our numerical results
demonstrate without any doubt that the shearing epicyclic motion is 
possible, and does decay in the presence of MHD turbulence.

A further theoretical expectation is as follows.  In an inviscid disc,
the epicyclic motion can decay by exciting inertial waves through a
parametric instability (Gammie, Goodman \& Ogilvie 2000, submitted).
In the optimal case, the signature of these waves is motion at
$30\degr$ to the vertical, while the wave vector is inclined at
$60\degr$ to the vertical.  The characteristic local growth rate of
the instability is
\begin{equation}
  \gamma = \frac{3\sqrt{3}}{16} 
\left|\frac{\partial u_x}{\partial z}\right|.
\end{equation}
This
instability can lead to a rapid damping of a warp, but may be somewhat
delicate as it relies on properties of the inertial-wave spectrum.  It
is important to determine whether it occurs in the presence of MHD
turbulence.

\begin{table}
\caption{Specification of numerical simulations:  the number of grid points
are given by $N_x \times N_y \times N_z$, and the amplitude of the initial
velocity perturbation is $u_0$.  Note that all Runs except Run 4 starts from
a snapshot of a previous simulation of turbulence in an accretion disc.
Run 4 starts from a state with no turbulent motions}
\begin{tabular}{rrr}
Run & $N_x \times N_y \times N_z$ & $u_0$ \\
0 & $31 \times 63 \times 63$ & 0.0\\
1 & $31 \times 63 \times 63$ & 0.011\\
1b & $63 \times 127 \times 127$ & 0.011\\
2 & $31 \times 63 \times 63$ & 0.095\\
3 & $63 \times 127 \times 127$ & 0.095\\
4 & $63 \times 127 \times 127$ & 0.095\\
\end{tabular}
\label{runs}
\end{table}

\section{Numerical simulations}
 
\subsection{Computational method}

We use the code by Nordlund \& Stein \shortcite{nordlund:stein} with the
modifications that were described by Brandenburg et al. 
\shortcite{brandenburg95}.  The code solves the MHD equations for $\ln 
\rho$, ${\bf u}$, $e$ and the vector potential ${\bf A}$, which gives the
magnetic field via ${\bf B} = {\bf \nabla \times A}$.  For the (radial)
azimuthal boundaries we use (sliding-) periodic boundary conditions.
The vertical boundaries are assumed to be impenetrable and stress-free.
Unlike our earlier studies, we now adopt perfectly conducting vertical
boundary conditions for the magnetic field.  Thus we have
\begin{equation}
  \frac{\partial u_x}{\partial z} = \frac{\partial u_y}{\partial z} =
u_z = 0,
\end{equation}
and 
\begin{equation}
  \frac{\partial B_x}{\partial z} = \frac{\partial B_y}{\partial z} =
B_z = 0.  
\end{equation}

We choose units such that $H = GM = 1$.
Density is normalized so that initially $\rho = 1$ at the
midplane, and we measure the magnetic field strength in velocity units, which
allows us to set $\mu_0 = 1$.  
The disc may be considered to be thin by the assumptions of our model, and the
results will thus not depend on the value of $R_0$.  We choose to set
$R_0 = 10$ in our units, which gives the orbital period $T_0 = 2\pi/\Omega_0 =
199$, and the
mean internal energy $e = 7.4\,10^{-4}$.  
The size of the box is
$L_x : L_y : L_z = 1 : 2\pi : 4$, where $x$ and $z$ vary between $\pm
\frac{1}{2}L_x$ and $\pm \frac{1}{2} L_z$, respectively, and $y$ goes from
0 to $L_y$.  The number of grid points is
$N_x \times N_y \times N_z$.
To stop the box from heating up during the simulation we introduce the 
cooling function
\begin{equation}
  Q = -\sigma_{\rm cool} \left(e -e_0\right),
\end{equation}
where $\sigma_{\rm cool}$ is the cooling rate, which typically corresponds to a
timescale of 1.5 orbital periods,
and $e_0$ is the internal energy
of an isothermal disc.

Although the Balbus-Hawley instability appears readily in a numerical 
simulation, experience has taught us that the initial conditions must
be chosen carefully in order for the instability to develop into 
sustained turbulence, unless the initial magnetic field has a net flux.
In particular, the initial field strength should be chosen such that
the Alfv\'en speed is close to the sound speed. Otherwise the field
becomes too weak before the dynamo effect sets in.  Experience has shown
that after about 50 orbital periods the simulations are independent of the
details of the initial conditions, which is typical of turbulence in
general.  To save time we started the simulations in this paper 
from a snapshot of a previous
simulation
\cite{brandenburg}.
The origin of this snapshot goes back to the simulations by Brandenburg et al. 
\shortcite{brandenburg95}.  In those simulations we started from a
magnetic field of the form $\hat {\bf z} B_0 \sin (2\pi x/L_x)$ (at that 
time we employed boundary conditions that constrained the magnetic field
to be vertical on the upper and lower boundaries).  $B_0$ was chosen such that
$\beta = 2\mu_0 p/B_0^2 = 100$ on average in the shearing box.  
A snapshot from an 
evolved stage of one of these simulations was later relaxed for about 55
orbital periods to fit the
perfectly conducting upper and lower boundaries of Brandenburg 
\shortcite{brandenburg}. 
For the purposes of this paper a snapshot from the new simulation was modified
in the following way.  For every horizontal layer in the snapshot
we subtract the mean horizontal velocity and
then add a net radial flow of the form
\begin{equation}
  u_x = u_0 \sin \left(\frac{\pi z}{L_z}\right).
\end{equation}
Two things that should be kept in mind here are that 
the previous simulations have been run long enough that the current snapshot
has lost its memory of its original initial conditions, 
and that in spite of all the 
modifications there is still no net magnetic flux in the shearing box.
The number of grid points and $u_0$
for the different runs are given in Table \ref{runs}.  $u_0$ should be compared
to the adiabatic sound speed which is 0.029.
We include one run, Run 0, in which we do not excite an epicyclic motion, as
a reference.

\begin{figure}
\epsfxsize=8.8cm
\epsfbox{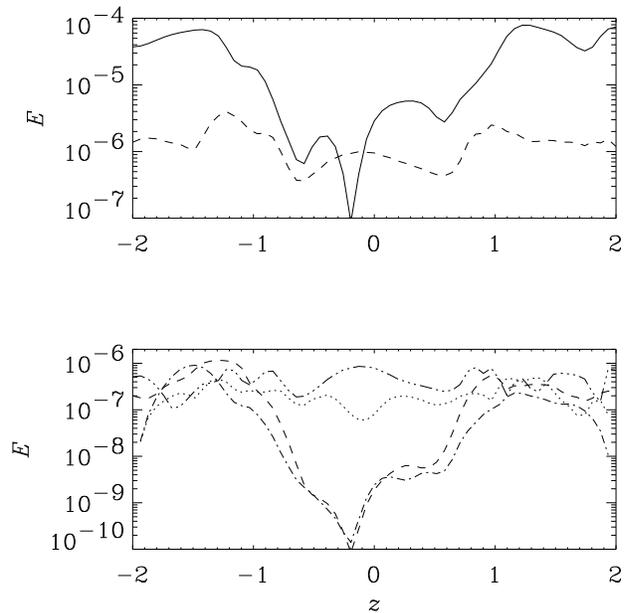}
\caption{Top:  The magnetic (solid line) and kinetic (dashed line) turbulent
energies as a function of $z$ at the beginning of Run 0.  The energies are
completely dominated by the contributions from the $y$-components of the 
magnetic field, and velocity, respectively.  Bottom:  $B_x^2/2$
(dashed line), $B_z^2/2$ (dot-dashed line), $\rho u_x^2/2$ 
(triple-dotted dashed line) and $\rho u_z^2/2$ (dotted line) as a function
of $z$}
\label{en_run0}
\end{figure}

\begin{figure}
\epsfxsize=8.8cm
\epsfbox{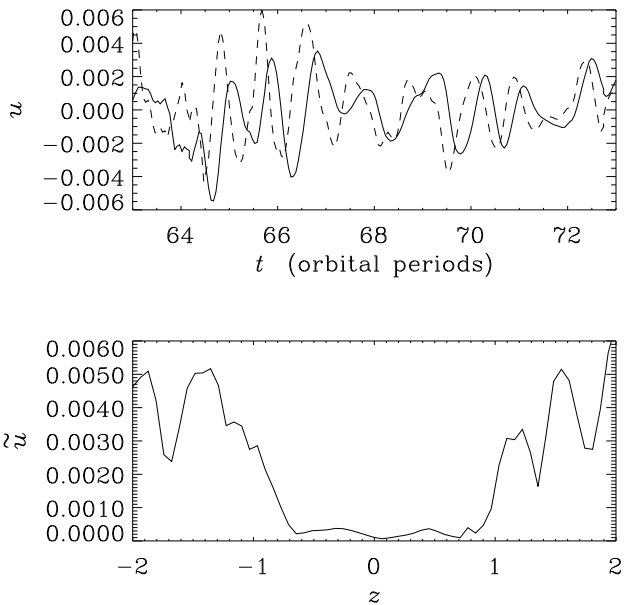}
\caption{Top:  $\langle u_x\rangle$ (solid line) and $2\langle u_y\rangle$
(dashed line) as a function of time at $z = 1.55$ for Run 0.  
The plot shows that 
epicyclic motions of amplitude $\sim 0.005$ lasting for a couple of 
orbits appear spontaneously in the disc.  Bottom:  $\tilde u$ as a function
of $z$ at $t = 66.7$ orbital periods}
\label{epic_run0}
\end{figure}

\begin{figure}
\epsfxsize=8.8cm
\epsfbox{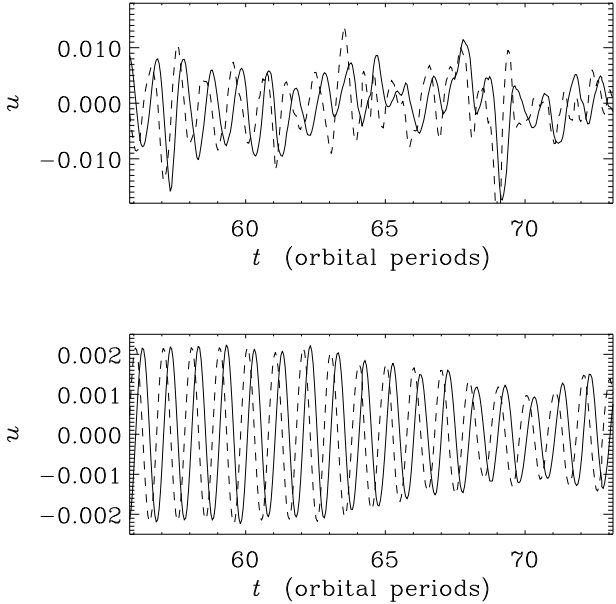}
\caption{$\langle u_x\rangle$ (solid line) and $2\langle u_y\rangle$
(dashed line) as a function of time at $z = 1.17$ (top) and $z = -0.25$ 
(bottom) of Run 1b.  
The damping timescale in the lower plot during the interval 63 to
68 orbital periods is 15.5 orbital periods}
\label{epic_run6b}
\end{figure}

\begin{figure}
\epsfxsize=8.8cm
\epsfbox{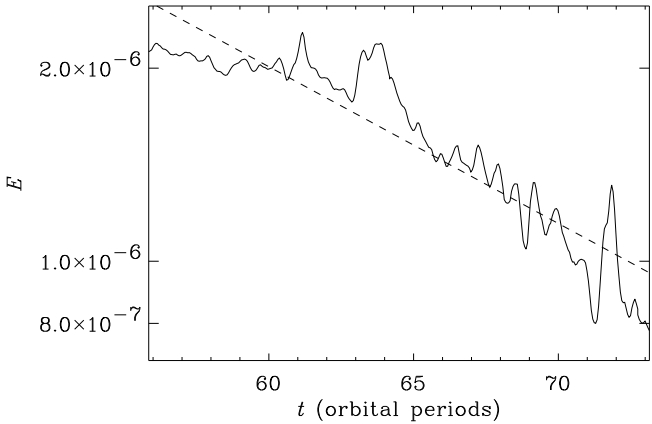}
\caption{$\langle E\rangle_{\rm V}$, 
the square of the epicyclic momentum, as a function of time
for Run 1b.  The dashed line is a fit with an exponential to $\langle E
\rangle_{\rm V}$.  The 
$e$-folding timescale of the exponential is 17.8 orbital periods}
\label{epic6b_en}
\end{figure}

\subsection{Results}

We start by looking at Run 0.  In this case we have not modified the 
velocity field, that is Run 0 represents the typical properties of the
turbulence.  The turbulence is fully developed at the beginning of our study
meaning that the shearing box does not remember its initial state any
longer.  We plot the
vertical distributions of the magnetic and turbulent kinetic energies in
Fig. \ref{en_run0}.  The kinetic energy is independent of the vertical 
coordinate $z$ while the magnetic energy has a minimum close to the 
midplane.  
It is a typical property of all our
simulations that the magnetic energy is up to an order of magnitude larger
than the turbulent kinetic energy, but the magnetic energy is still
an order of magnitude smaller than the internal energy, whose mean density as
stated above is $7.4\,10^{-4}$.  A property of both the magnetic and turbulent
kinetic energies is that they are completely dominated by the contributions
from the azimuthal components of the magnetic field and turbulent velocity,
respectively.  The other components of the turbulent velocity and magnetic
field make about equal contributions to the energy density far from the 
midplane of the disc.  

The turbulence generates
epicyclic motions with amplitudes $\sim 0.004$
lasting for a couple of orbital periods (Fig. \ref{epic_run0}).  
The epicyclic velocity amplitude $\tilde u$ (see Eq. \ref{epic_amp}) is peaked 
towards the surfaces of the accretion disc (cf. Fig. \ref{epic_run0}, bottom).
Overall these motions complicate the analysis of the
rest of our numerical simulations, and force us to excite motions with
amplitudes significantly larger than that of the motions produced by 
the turbulence.

Figure \ref{epic_run6b} shows the mean horizontal motion of
Run 1b.  At the start of the simulation we added a radial velocity of
amplitude 0.011.
Far from the midplane the imposed epicyclic motion is comparable to that 
generated by the turbulence, and it becomes virtually impossible to identify
a phase of exponential decay.  Closer to the midplane the imposed
epicyclic motion is initially unaffected by the turbulence.  After seven
or eight orbital periods a damping sets in, but the net damping before the
turbulence starts to reinforce the epicyclic motion is less than a factor of
two.  By looking at the
volume average of the square of the epicyclic 
momentum, $\langle E\rangle_{\rm V}$, (Fig. \ref{epic6b_en}) we find
a clear trend.  The epicyclic motion is obviously damped,
and by fitting an exponential we obtain an $e$-folding timescale of 17.8
orbital periods for $\langle E\rangle_{\rm V}$, 
corresponding to a timescale of 35.6 orbital periods for
the momentum, but the damping is not described accurately by an exponential.
Initially the damping is much slower than the fitted exponential, while
at the end  it is faster.

\begin{figure}
\epsfxsize=8.8cm
\epsfbox{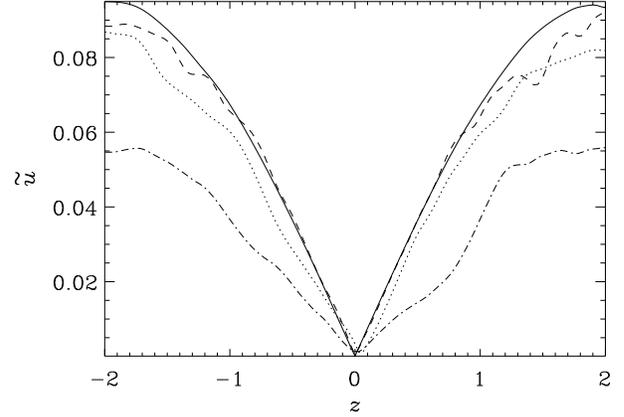}
\caption{$\tilde u$ 
as a function of $z$ at $t = 55.8 \,T_0$ (solid line), $57.1 \,T_0$ (dashed
line), $58.4 \,T_0$ (dotted line), and $60.9 \,T_0$ (dot-dashed line) for Run 3}
\label{epic4_big}
\end{figure}

\begin{figure}
\epsfxsize=8.8cm
\epsfbox{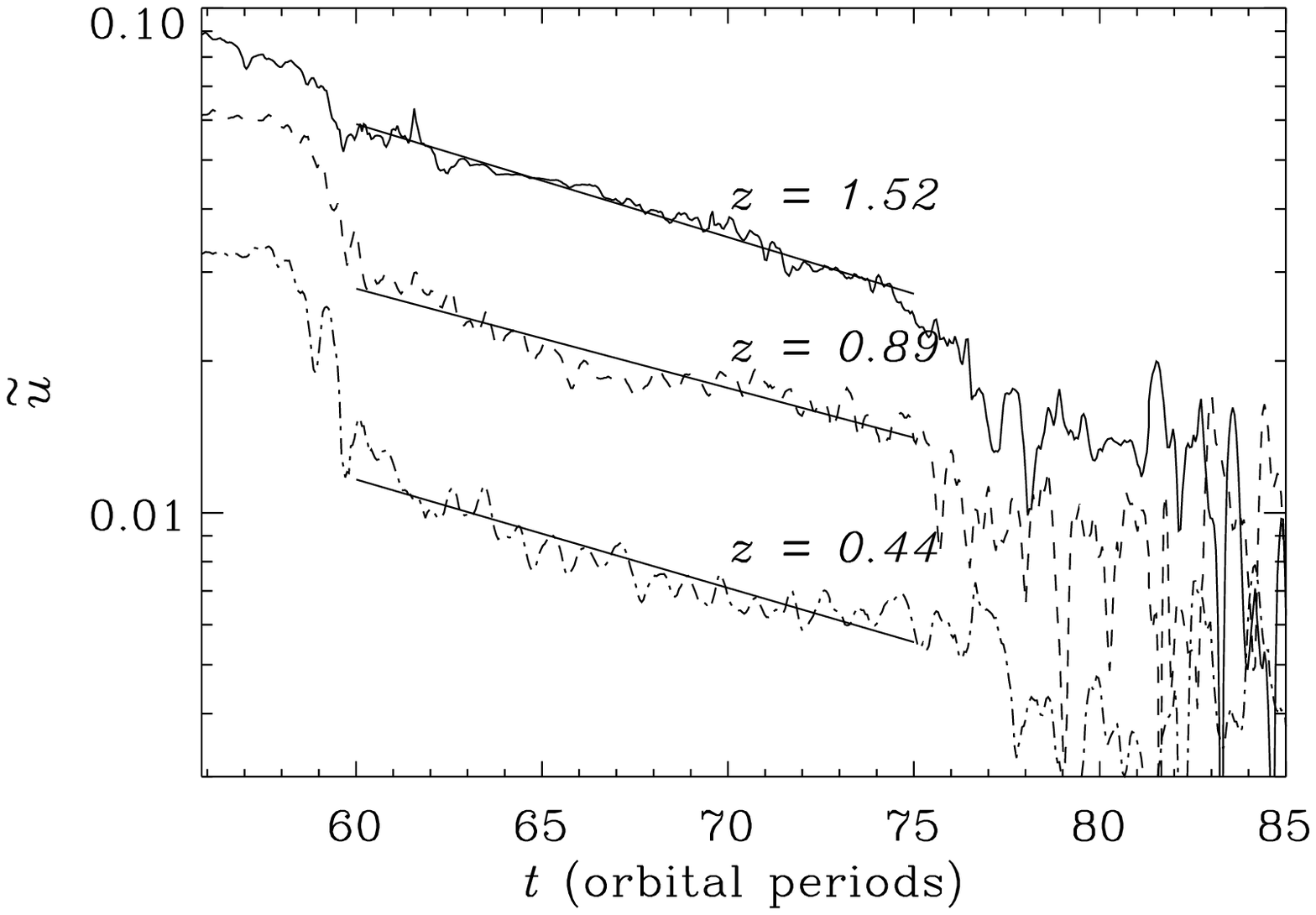}
\caption{The amplitude of the epicyclic motion in Run 3, $\tilde u$, 
as a function of $t$ on
three horizontal planes:  $z = 1.52$ (solid line), $z = 0.89$ (dashed line) and 
$z = 0.44$ (dot-dashed line).  The straight lines are exponential functions
that have been fitted for the interval $60 T_0 < t < 75 T_0$.  The $e$-folding
timescales of the exponentials starting from the top are $19.4 T_0$, $22.1 T_0$
and $20.2 T_0$, respectively.  
The epicyclic motion was added to the box at $t =
55.7 \,T_0$}
\label{motion_1}
\end{figure}

\begin{figure}
\epsfxsize=8.8cm
\epsfbox{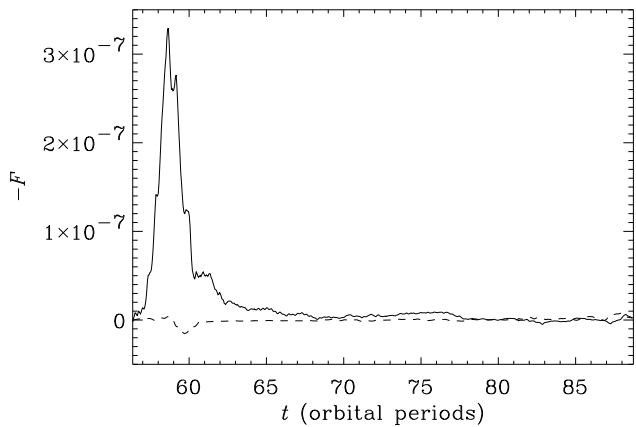}
\caption{$-\overline{\langle F_u\rangle_{\rm V}}$ (solid line) and 
$-\overline{\langle F_B\rangle_{\rm V}}$ (dashed line) as a function of time
for Run 3}
\label{force}
\end{figure}

The analysis becomes more straightforward in Run 3, where we
excite a motion of amplitude 0.095. There is then a
sufficient dynamic range between the epicyclic and turbulent
motions.
We show the vertical variation of $\tilde u$ at four
different times in Fig. \ref{epic4_big}, and plot it as a function of time
at three different heights;
$z = 1.52$, $z = 0.89$ and $z = 0.44$, in Fig.
\ref{motion_1}.  
Figure \ref{epic4_big} shows that the damping sets in first at the surfaces,
while for $|z| < 1$ there is essentially no damping during the first two 
orbital periods (Fig. \ref{motion_1}).
There is then a brief period of rapid damping between $t = 58 \,T_0$ and 
$t = 60 \,T_0$ throughout the box, especially for small $|z|$ where $\tilde u$
may drop by a factor 2.  This is followed by a period of exponential decay, but
after $t = 75 \,T_0$ 
it becomes difficult to follow the epicyclic motion, as the influence 
of the random turbulence on $\tilde u$ becomes significant, in particular 
close to the midplane, where $\tilde u$ is anyway small.
We estimate the damping time, $\tau = (\mbox{d}\ln \tilde u/\mbox{d}t)^{-1}$
by fitting exponentials to $\tilde u$ in the interval $60 T_0 < t < 75 T_0$.
Averaged over the box we get $\tau = 25 \pm 8 T_0$, which corresponds to
$\alpha_{\rm v} = 0.006\pm 0.002$ according to Eq. (\ref{alpha-v}).  

We may determine the influence of the Maxwell and Reynolds stresses on the
shear flow by plotting $F_u$ and $F_B$ as functions of time (Fig. \ref{force}).
The sharp peak in $F_u$ coincides with the parametric decay of the 
epicyclic motion to inertial waves (see Sect. \ref{param}).  As this is a
hydrodynamic process it is not surprising that $F_u > F_B$, but it was not
expected that $F_u$ would remain the dominating effect at later times,
since at the same time it is the Maxwell stress that is driving the radial
accretion flow.

\begin{figure}
\epsfxsize=8.8cm
\epsfbox{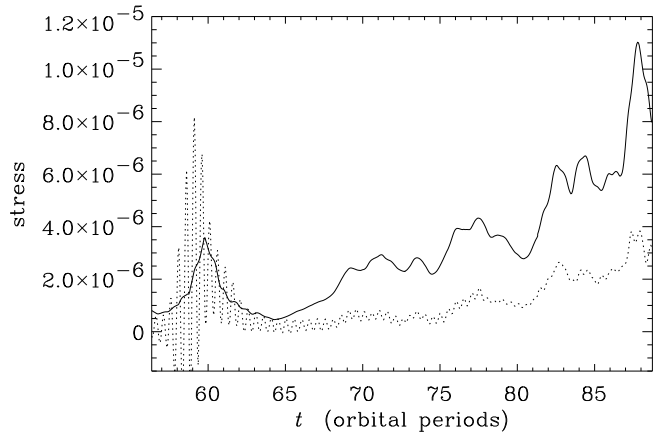}
\caption{
$-\overline{\langle B_x B_y\rangle_{\rm V}}$ (solid line) and 
$\overline{\langle\rho u_x u_y \rangle_{\rm V}}$ of
Run 3}
\label{mstress_big}
\end{figure}

\begin{figure}
\epsfxsize=8.8cm
\epsfbox{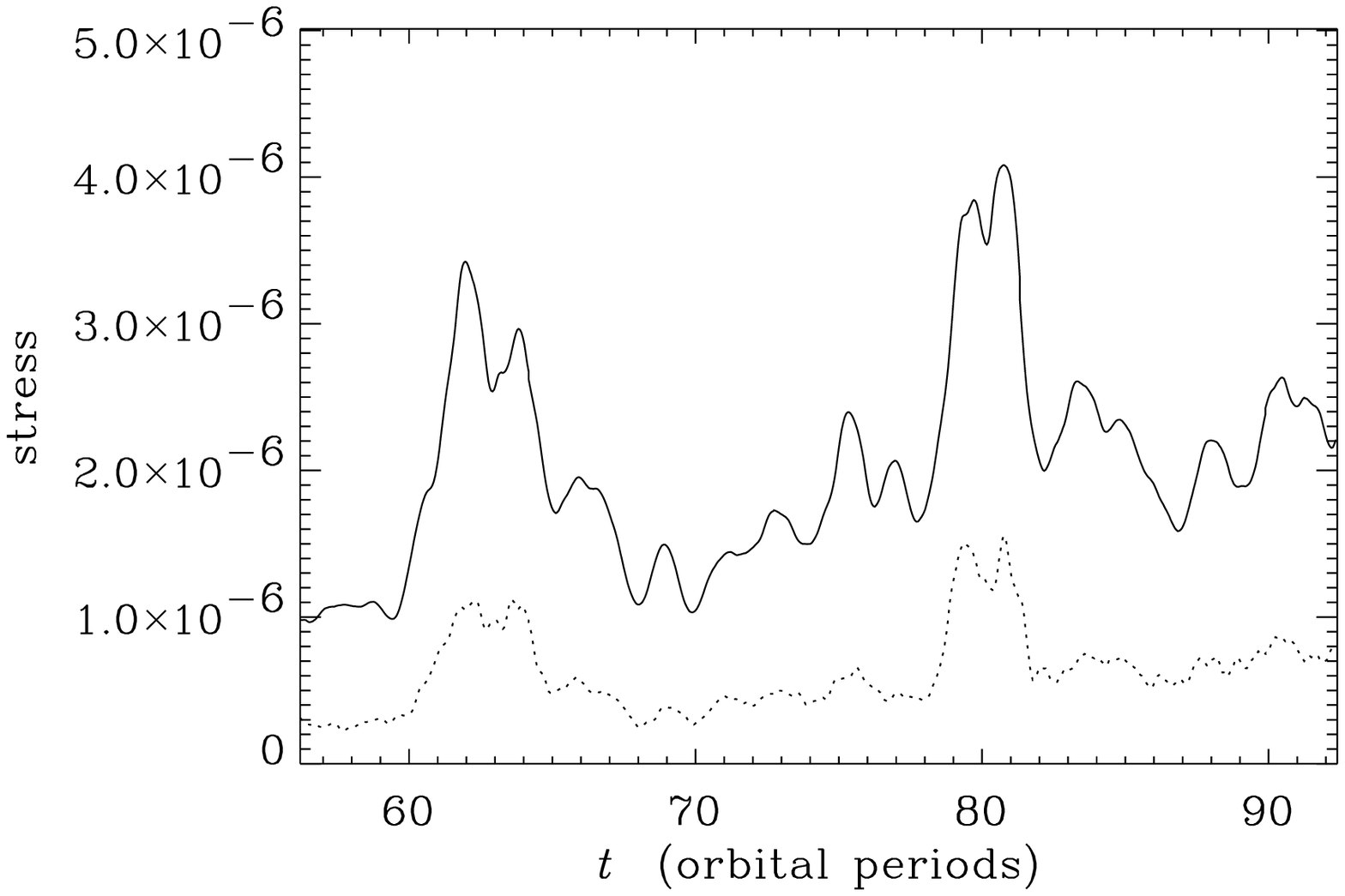}
\caption{
$-\overline{\langle B_x B_y\rangle_{\rm V}}$ (solid line) and 
$\overline{\langle\rho u_x u_y \rangle_{\rm V}}$ of
Run 0}
\label{mstress}
\end{figure}

\begin{figure}
\epsfxsize=8.8cm
\epsfbox{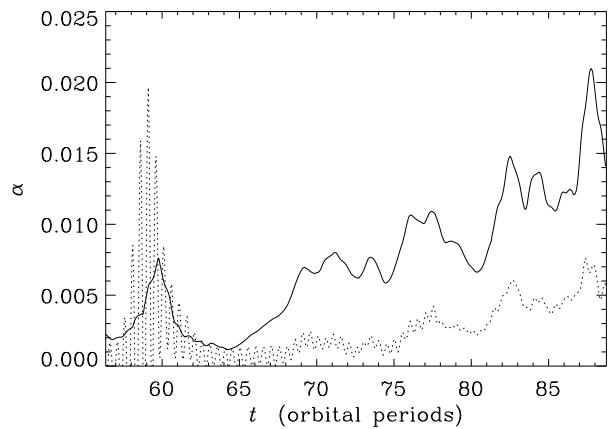}
\caption{$\alpha_{\rm h}$ is calculated by dividing
$-\overline{\langle B_x B_y\rangle_{\rm V}}$
(solid line) and $\overline{\langle\rho u_x u_y\rangle_{\rm V}}$ (dashed line) 
of Run 3 with
$\frac{3}{2}\overline{\langle p\rangle_{\rm V}}$}
\label{alpha}
\end{figure}

The accretion itself is driven by the $\langle \rho u_x u_y - 
B_x B_y\rangle$-stresses.
We plot the moving time averages of the vertical average of the 
accretion-driving 
stress of Run 3 in Fig. \ref{mstress_big}, and as
a comparison that of Run 0 in Fig. \ref{mstress}.
In the beginning of Run 3 the Reynolds stress 
is modulated on a timescale of half the
orbital period.  This modulation is an artifact
of the damping of the epicyclic motion and dies out with time.
The  Maxwell stress becomes significantly stronger
than the Reynolds stress once the epicyclic motion has vanished.  Later
the stresses
vary in phase with each other, like they do all the time in Run 0.
The main difference between Run 0 and the end of Run 3 is that the stresses
are 2-3 times larger in Run 3.  
We calculate $\alpha_{\rm h}$ for Run 3
by dividing the stresses by $\frac{3}{2}\overline{\langle p\rangle_{\rm V}}$
(Fig. \ref{alpha}).
One should take note that the simulations in this paper are too short 
to derive $\alpha_{\rm h}$ with high 
statistical significance (cf. Brandenburg et al. 1995), but our results do show
that $\alpha_{\rm h}$ varies in phase with the stress.  In other words the
pressure variations are smaller (the pressure increases by 50\% during the
course of the simulation) than the stress variations, which is not what we
expect from Eq. (\ref{alpha-h}).  The lack of a correlation between the stress
and the pressure is even more evident in Run 0, in which the pressure never 
varies with more than 5\%.

\begin{figure}
\epsfxsize=8.8cm
\epsfbox{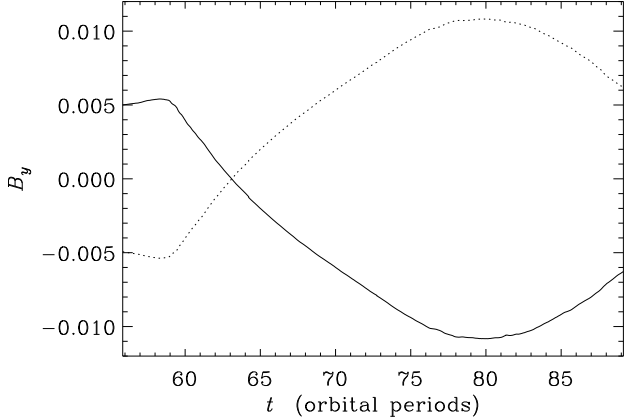}
\caption{$B_y$ averaged over the upper half (solid line) and
lower half (dotted line) of the box of Run 3}
\label{by}
\end{figure}

In our previous work \cite{brandenburg95} we found that the toroidal
magnetic flux reversed its direction about every 30 orbital periods.  With
the perfect conductor boundary conditions that we have assumed in this paper
such field reversals are not allowed, since the boundary conditions conserve
the toroidal magnetic flux (e.g. Brandenburg 1999).  
In all of our simulations we made sure that the toroidal magnetic flux was
0.
However the azimuthal magnetic
field organized itself in such a way that it is largely
antisymmetric with respect to the
midplane.  That is, considering separately the upper and lower halves of
the box, we may still 
find significant toroidal fluxes, but with opposite directions.  
These fluxes may still be reversed by the turbulent dynamo as we found
in Run 3 (Fig. \ref{by}).

\subsection{Dependence on $u_0$ and on the resolution}

There is some evidence from comparing Run 1b and Run 3 
that the damping timescale decreases as $u_0$ is
increased, but it
is more difficult to study the damping of the epicyclic motion for a 
smaller $u_0$, as the turbulence may excite epicyclic motions on its own.
The randomly excited motions swamp the epicyclic motion that we are 
studying.  The quantitative results from Runs 1
and 1b are therefore more uncertain.
In other respects the turbulent stresses of Runs 1 and 1b are more 
similar to those
of Run 0 than to those of Run 3.

On the other hand there are significant differences 
between Runs 2 and 3, which differ only in terms of the grid 
resolution.  The magnetic field decays rapidly in Run 2, and only the toroidal
field recovers towards the end of the simulation.  In the absence of a poloidal
magnetic field there is no magnetic stress, and therefore the disc cannot
extract energy from the shear flow.  
Consequently there is no turbulent heating in
Run 2, and the disc settles down to an isothermal state.  
Apparently the turbulence is killed by the numerical diffusion in Run 2.
This demonstrates that the minimal resolution which is required in the
simulation depends on the amplitude $\tilde u_0$.  A simulation 
with an imposed velocity $\tilde u_0$
with an amplitude significantly larger than that of the turbulence requires 
a finer resolution than a simulation of undisturbed turbulence.  Run 2
failed because the imposed velocity field in combination with
the limited resolution generated a numerical diffusion large enough to
kill the turbulence.
This was not a problem in Run 1, where the amplitude $\tilde
u_0$ is much smaller.

\begin{figure}
\epsfxsize=8.8cm
\epsfbox{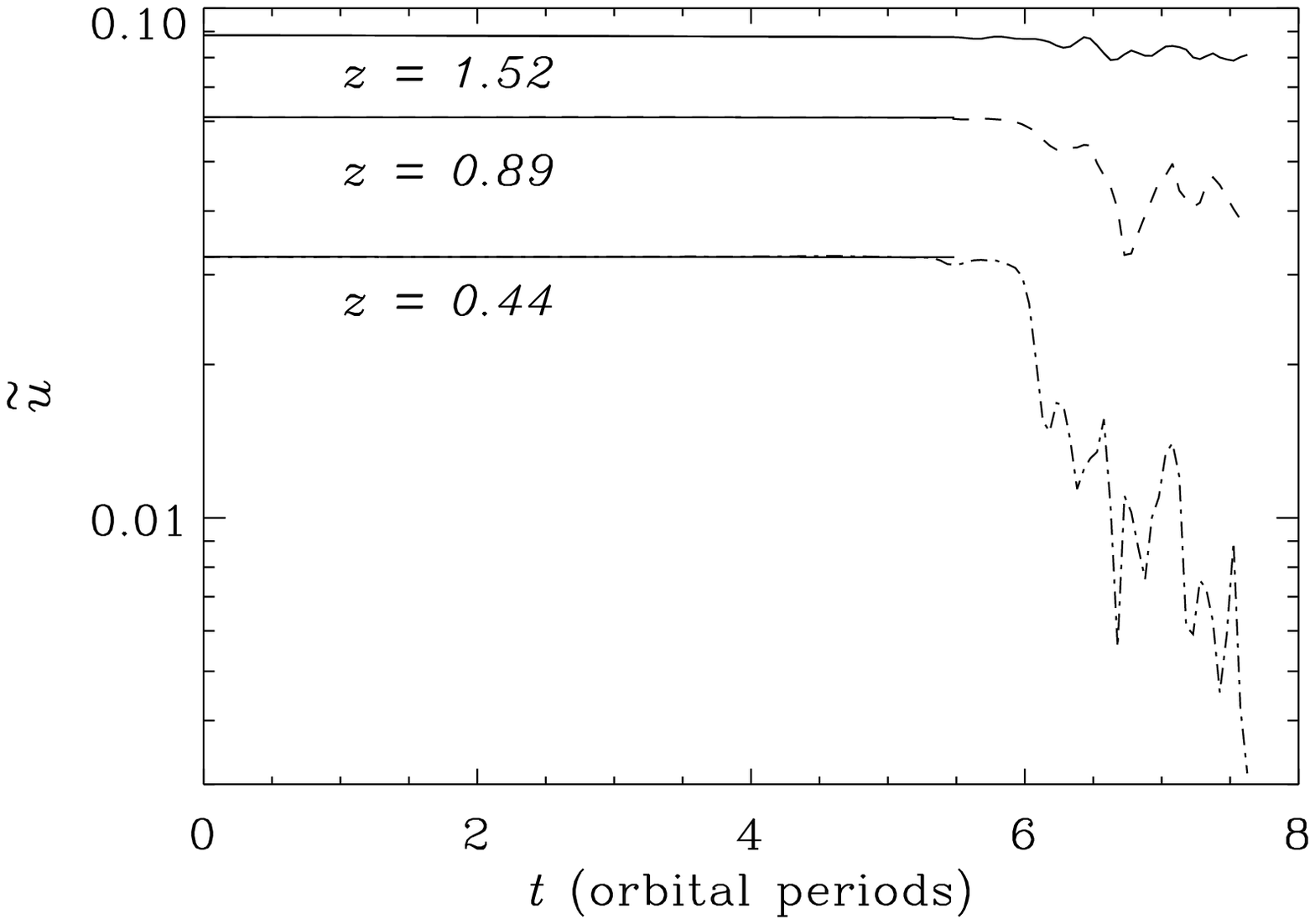}
\caption{The amplitude of the epicyclic motion of Run 4, $\tilde u$, as a 
function of $t$ on three horizontal planes:  $z = 1.52$ (solid line), 
$z = 0.89$ (dashed line), and $z = 0.44$ (dot-dashed line).  The straight
lines are exponential functions that have been fitted for the interval
$0 < t < 5.5 T_0$.  The $e$-folding timescales of the exponentials
starting from the top are $640 T_0$, $4\,700 T_0$ and $2\,800 T_0$, 
respectively}
\label{epic_comp}
\end{figure}

To check the influence of numerical diffusion on Run 3 we imposed the
same velocity profile as in Run 3 on a shearing box without any turbulence
(Run 4).
We plot the evolution of $\tilde u$ in Fig. \ref{epic_comp}.  We
fitted exponential functions to $\tilde u$ for the interval 0 to 5.5
orbital periods.  The shortest $e$-folding time we obtained over this
interval was 640 orbital periods for $z = 1.52$.  Closer to the midplane the
$e$-folding timescale was several thousand orbital periods.  These 
numbers are indicative of the effect of numerical diffusion in the 
simulations.  The strong damping at $t = 6$ orbital periods is {\em not} due
to numerical diffusion, rather it is caused by a {\em physical} parametric
instability (see Sect. \ref{param})
that, in this case, has been triggered by numerical noise.

\section{Discussion}

For the purposes of the following discussion we now summarize our
main results:
\begin{itemize}

\item  In Runs 3 and 4 the initial damping of the epicyclic motion is
caused by its parametric decay to inertial waves (see Sect. \ref{param}).

\item  Apart from this, the epicyclic motion experiences 
approximately exponential damping through interaction with the turbulence.  
The $e$-folding timescale is about
25 orbital periods, which may be interpreted as $\alpha_{\rm v} = 0.006$.

\item  $\alpha_{\rm h}$, which describes the accretion-driving stress
$\langle \rho u_x u_y - B_x B_y\rangle$ is of comparable size.

\end{itemize}

\subsection{Parametric decay to inertial waves}

\label{param}

\begin{figure}
\epsfxsize=8.8cm
\epsfbox{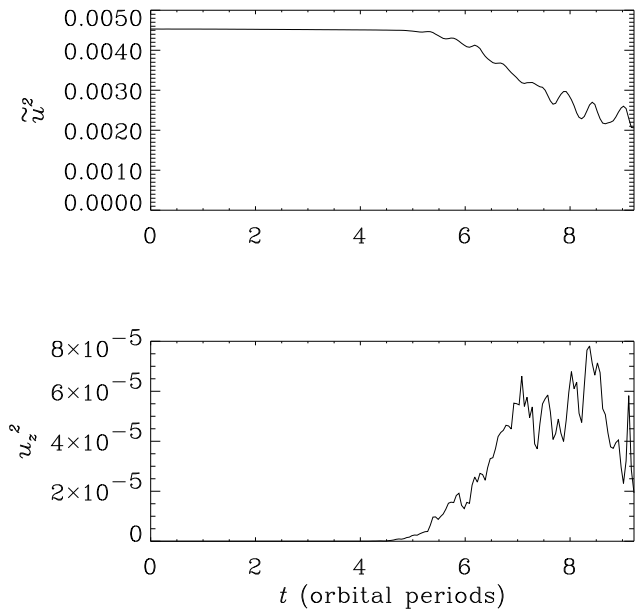}
\caption{$\langle\tilde u^2\rangle_{\rm V}$ (top) and 
$\langle u_z^2\rangle_{\rm V}$ (bottom)
for a two-dimensional simulation of parametric decay.  The decay of the
epicyclic motion excites a vertical motion}
\label{parametric}
\end{figure}

\begin{figure}
\epsfxsize=8.8cm
\epsfbox{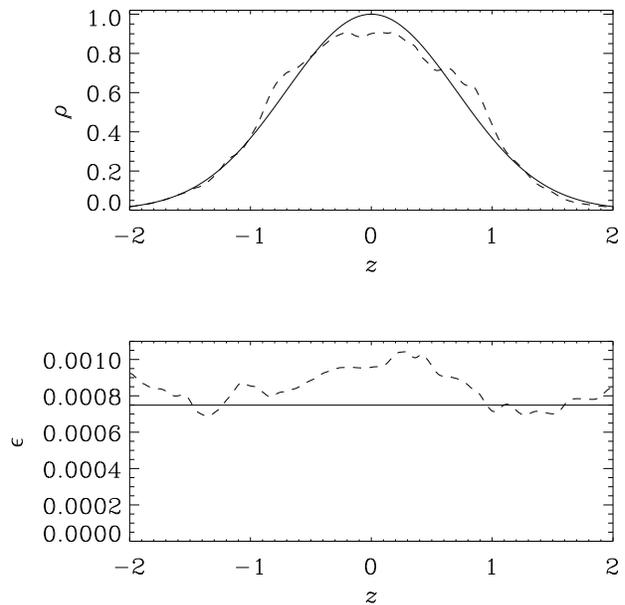}
\caption{The density $\langle \rho\rangle$ (top) and internal
energy $\langle e\rangle$ (bottom) as functions of $z$
at the start of the two-dimensional simulation (solid line) and after 
8.97 orbital periods (dashed line)}
\label{rho_2d}
\end{figure}

Gammie et al. \shortcite{gammie98} predicted the occurrence of a
parametric instability in epicyclic shear flows.  The shear flow should 
excite pairs of inertial waves that propagate at roughly a $30\degr$ angle to 
the vertical, and involve vertical as well as horizontal motions.  
To elucidate the dynamics of the parametric instability we make use of a
two-dimensional hydrodynamic simulation of an epicyclic shear flow.
Our two-dimensional $xz$-plane has the same extension as in the previous 
three-dimensional simulations, but we are now using $128\times 255$ grid 
points.  The initial state is a stratified Keplerian 
accretion disc to which we have added a radial motion with amplitude
0.095, and a small random perturbation of the pressure.  
Initially the amplitude of the epicyclic motion, $\tilde u$, is
constant, but the damping sets in suddenly after 5 orbital periods (Fig. 
\ref{parametric}).  Over four orbital periods the amplitude of the epicyclic
motion decreases by 30\%, after which the damping becomes weaker again.  This
is clearly not an exponential damping.
At the same time as the epicyclic motion is damped
the vertical velocity starts to grow.
The inertial waves themselves are not capable of transporting mass, but the
parametric decay heats the central part of the disc, and the corresponding
pressure increase pushes matter away from the midplane (Fig. \ref{rho_2d}).

\begin{figure}
\epsfxsize=8.8cm
\epsfbox{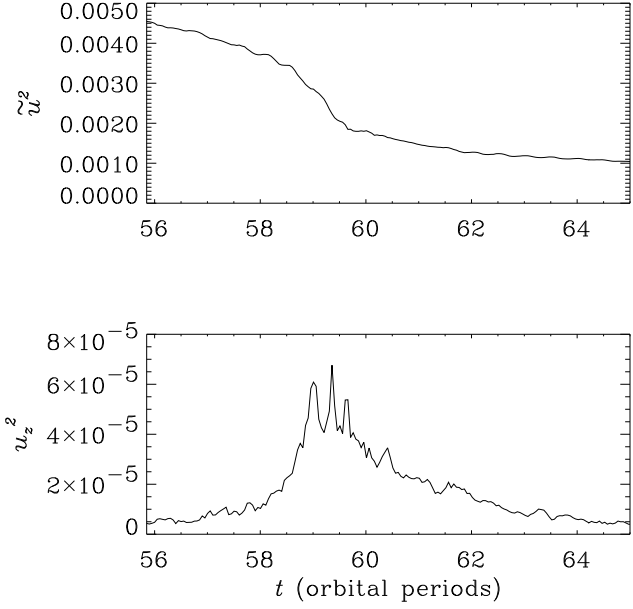}
\caption{$\langle\tilde u^2\rangle_{\rm V}$ (top) and 
$\langle u_z^2\rangle_{\rm V}$ (bottom)
for Run 3}
\label{param_real}
\end{figure}

\begin{figure}
\epsfxsize=8.8cm
\epsfbox{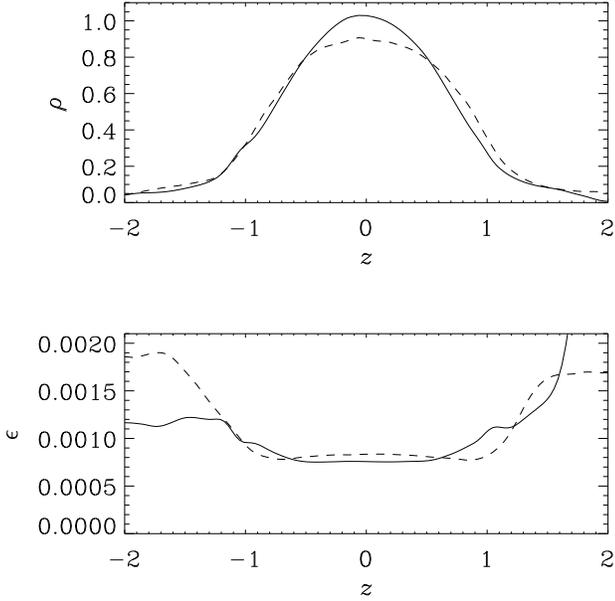}
\caption{The density $\langle \rho\rangle$ (top) and internal energy 
$\langle e\rangle$
(bottom) as functions of $z$
for Run 3 at $t = 55.8$ (solid line) and $t = 60.9$ (dashed line) orbital
periods}
\label{rho_real}
\end{figure}

The same parametric decay may be found in Run 4 (Fig. \ref{epic_comp}), 
which is done on the same grid as Run 3, but starting from a laminar state
with only the epicyclic shear flow.
In Run 3 the parametric decay occurs between orbits 58 and 60 
(Fig. \ref{param_real}),
but it is followed by an exponential damping caused by the turbulence
itself (Fig. \ref{motion_1}).  The damping rate for Run 3 as estimated in
Sect. 3.2 is based on a time interval after the end of the
episode of parametric decay, and therefore represents the turbulent damping
rate.
Also, in Runs 3 and 4 we find an enhanced heating of the central regions of the
disc and a resulting density reduction (Fig. \ref{rho_real}).

\subsection{Application to warped accretion discs}

We now investigate the implications of our results for the large-scale
dynamics of a warped accretion disc.
In linear theory we may estimate the
amplitude of the epicyclic motion as
(cf. Eq. \ref{epic_vel})
\begin{equation}
  \frac{u_0}{c_{\rm s}} \sim \frac{A}{\alpha_{\rm v}},
\end{equation}
where $A$ is the
dimensionless amplitude of the warp.
This estimate is valid for a thin and sufficiently viscous disc, that is for
$H/r \la \alpha_{\rm v} \ll 1$ \cite{papaloizou:pringle}.  
For observable warps in which $A$ exceeds the aspect ratio of
the disc, the epicyclic velocities are comparable to, or greater than, the
sound speed, as is the case in our numerical simulations.  
Based on the high velocities one might expect shocks to appear in the 
simulations, and shocks do appear in some global simulations that allow
horizontal gradients in the epicyclic velocity (cf. Nelson \& Papaloizou 1999).
However, because of the local nature
of our model we do not find any such gradients or shocks.

The large-scale dynamics of a warped disc has been formulated by Pringle
(1992) in terms of two effective kinematic viscosity coefficients: $\nu_1$
describes the radial transport of the component of the angular momentum
vector parallel to the tilt vector, while $\nu_2$ describes the transport
of the perpendicular components.  The relation between $\nu_1$ and $\nu_2$
and $\alpha$ is non-trivial but has been explained by Papaloizou \&
Pringle (1983) and Ogilvie (1999).  Recall that $\alpha_{\rm h}$ and 
$\alpha_{\rm v}$
are effective viscosity parameters that represent the transport of
momentum by turbulent motions (and magnetic fields) on scales small
compared to $H$.  Now $\nu_1\propto\alpha_{\rm h}$ as expected, because the
parallel component of angular momentum is transported mainly by these
small-scale motions and magnetic fields.  However
$\nu_2\propto\alpha_{\rm v}^{-1}$ instead of the intuitive result
$\nu_2\propto\alpha_{\rm v}$.  (Here we assume $H/r\la\alpha_{\rm v}\ll1$.)  
This is
because the perpendicular components of angular momentum are transported
mainly by the systematic epicyclic motions induced by the warp, and these
are proportional to $\alpha_{\rm v}^{-1}$ as explained in Section 2.3.  In
effect, $\nu_2$ is an effective viscosity coefficient removed by an
additional level from the turbulent scales.  If we generalize the linear
theory of Papaloizou \& Pringle (1983) to allow for an anisotropic
small-scale effective viscosity, we obtain
$\nu_2/\nu_1=1/(2\alpha_{\rm h}\alpha_{\rm v})$.

The condition for a warp to appear in the accretion disc is set by the
balance between the torque that is exciting the warp and the viscous
torque, described by $\nu_2$, that is flattening the disc.  The
warp-exciting torque may for instance be a radiation torque from the
central radiation source.  Assuming that accretion is responsible for
all the radiation, the radiation torque will depend on the viscosity
$\nu_1$.  The criterion for the warp to appear will then depend on
the ratio of viscosities $\eta = \nu_2/\nu_1$.  Pringle 
\shortcite{pringle} showed that an irradiation-driven warp will appear
at radii
\begin{equation}
  r \ga \left(\frac{2\sqrt{2}\pi \eta}{\epsilon}\right)^2 R_{\rm Sch},
\end{equation}
where $R_{\rm Sch}$ is the Schwarzschild radius, 
and $\epsilon = L/\dot Mc^2$ is the efficiency of
the accretion process. 
We have shown that 
$\alpha_{\rm h} \approx \alpha_{\rm v} \ll 1$.
However, we emphasize again that this does {\it not} imply that $\eta \approx
1$; on the contrary, we estimate that $\eta \approx 1/(2\alpha_{\rm h}
\alpha_{\rm v}) \gg 1$.  
The high value 
for $\eta$ will make it difficult for a warp to appear unless
the radiation torque can be amplified by an additional physical mechanism.
One way to produce a stronger torque is if the irradiation is driving an
outflow from the disc (cf. Schandl \& Meyer 1994).

A similar damping mechanism may affect waves excited
by Lense-Thirring precession in the inner part of the accretion disc around
a spinning black hole.  Numerical calculations by 
Markovic \& Lamb \shortcite{markovic:lamb} and Armitage \& Natarajan
\shortcite{armitage:natarajan} show that these waves are damped rapidly unless
$\nu_2 \ll \nu_1$, which we find is not the case.  (However, we
note that the resonant enhancement of $\nu_2$ will be reduced near the
innermost stable circular orbit, because the epicyclic frequency
deviates substantially from the orbital frequency).  Likewise a high value
of $\nu_2$ will lead to a rapid alignment of the angular momentum vectors
of a black hole and its surrounding accretion disc
(cf. Natarajan \& Pringle 1998).

Some caution should be exercised when interpreting the results of this 
paper.  Although
the shearing box simulations in general have been successful in demonstrating
the appearance of turbulence with the right properties for driving 
accretion, they are in general producing uncomfortably low values of
$\alpha_{\rm h}$ to describe for instance outbursting dwarf nova discs
(e.g. Cannizzo, Wheeler \& Polidan 1986).  An underestimate of $\alpha_{\rm h}$
and $\alpha_{\rm v}$
would lead to an overestimate of $\eta$.
In addition the parametric instability leads to an enhanced damping of
the epicyclic motion.  If this effect had been sustained it would have resulted
in a smaller value for $\eta$.  In a warped accretion disc the epicyclic
motion is driven by the pressure gradients, which may maintain the velocities
at a sufficient level for the parametric instability to operate continuously.
To address this question we intend to study forced oscillations in a future 
paper.

\begin{figure}
\epsfxsize=8.8cm
\epsfbox{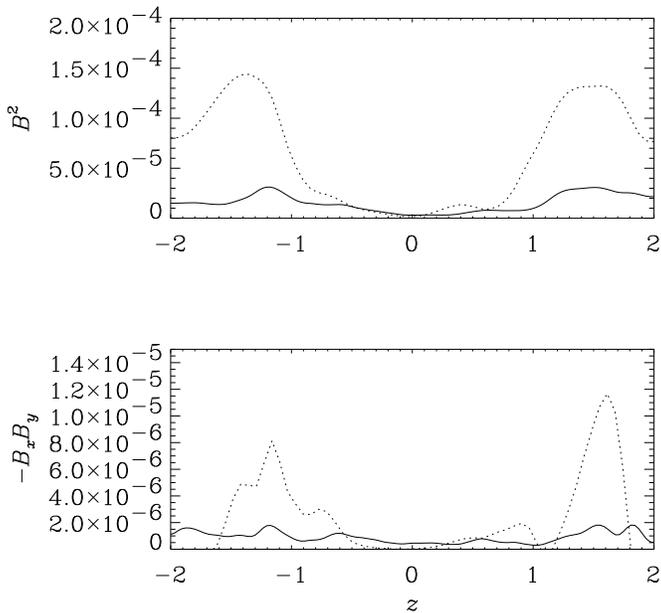}
\caption{The upper plot shows $\overline{\langle B^2\rangle}$, the average
of the magnetic energy, as a function of $z$ for Runs 3 at $66.3 T_0$
(solid line) and 0 at $66.1 T_0$
(dashed line).  The lower plot shows $-\overline{\langle B_x B_y\rangle}$
plotted the same way}
\label{bstress}
\end{figure}

\subsection{The vertical structure of the accretion disc}

Our previous simulations of turbulence in a Keplerian shearing box have shown
that the turbulent $xy$-stresses are approximately
constant with height \cite{kyoto}
rather than proportional to the pressure as may have been expected
from the 
$\alpha$-prescription \cite{shakura:sunyaev}.  We modified the
vertical boundary conditions for this paper and added 
an epicyclic motion.
The $\langle B_x B_y\rangle$-stress is still approximately
independent of $z$ or even increasing
with $|z|$ for $|z| < H$ though, 
while at larger $|z|$ we may see the effects of the
boundary conditions (Fig. \ref{bstress}).
The effect of the epicyclic motion is seemingly to limit
the $\langle B_x B_y\rangle$-stress in the surface layers to its value in
the interior of the disc.
The fact that the stresses decrease more slowly with $¦z¦$ than the density 
results in a strong heating of the surface layers.

\section{Conclusions}

In this paper we have studied how the turbulence in an accretion disc will
damp an epicyclic motion, whose amplitude depends on the vertical 
coordinate $z$ in the accretion disc.  Such a motion could be set up
by a warp in the accretion disc \cite{papaloizou:pringle}.  We find that
the typical damping timescale of the epicyclic motion is about 25 
orbital periods, which corresponds to $\alpha_{\rm v} = 0.006$.  This
value is comparable to the traditional estimate of $\alpha_{\rm h}$ 
that one gets from comparing
the $\langle \rho u_x u_y - B_x B_y\rangle$-stress with the pressure.
Both alphas are of the order of 0.01, which implies that the timescale
for damping a warp in the accretion disc is much shorter than the usual 
viscous timescale.
That the two alphas are within a factor of a few of each other is surprising,
since the damping of the epicyclic motion may be attributed to the Reynolds 
stresses,
while the accretion is mostly driven by the Maxwell stress.

However, not all of the damping can be described as a simple viscous damping.
At high amplitudes the epicyclic motion may also decay parametrically to 
inertial waves.  The damping is much more efficient in the presence of
this mechanism.

\section*{Acknowledgments}
UT was supported by an EU post-doctoral fellowship in Cambridge and
is supported by the Natural Sciences Research Council (NFR) in 
Gothenburg.  Computer
resources from the National Supercomputer Centre at Link\"oping University are
gratefully acknowledged.  
GIO is supported by the European Commission through the TMR network
`Accretion on to Black Holes, Compact Stars and Protostars' (contract 
number ERBFMRX-CT98-0195).
This work was supported in part by the Danish National
Research Foundation through its establishment of the Theoretical Astrophysics
Center (\AA N).  RFS is supported by NASA grant NAG5-4031.

\end{document}